# Energy, Momentum, and Force in Classical Electrodynamics: Application to Negative-index Media

Masud Mansuripur[†] and Armis R. Zakharian[‡]

[†]College of Optical Sciences, The University of Arizona, Tucson, Arizona 85721
[‡]Corning Incorporated, Science and Technology Division, Corning, New York 14831

**Abstract**. The classical theory of electromagnetism is based on Maxwell's macroscopic equations, an energy postulate, a momentum postulate, and a generalized form of the Lorentz law of force. These seven postulates constitute the foundation of a complete and consistent theory, thus eliminating the need for physical models of polarization $\boldsymbol{P}$ and magnetization $\boldsymbol{M}$ – these being the distinguishing features of Maxwell's macroscopic equations. In the proposed formulation, $\boldsymbol{P}(\boldsymbol{r},t)$ and $\boldsymbol{M}(\boldsymbol{r},t)$ are arbitrary functions of space and time, their physical properties being embedded in the seven postulates of the theory. The postulates are self-consistent, comply with special relativity, and satisfy the laws of conservation of energy, linear momentum, and angular momentum. The Abraham momentum density $\boldsymbol{p}_{\text{EM}}(\boldsymbol{r},t) = \boldsymbol{E}(\boldsymbol{r},t) \times \boldsymbol{H}(\boldsymbol{r},t)/c^2$ emerges as the universal electromagnetic momentum that does not depend on whether the field is propagating or evanescent, and whether or not the host media are homogeneous, transparent, isotropic, linear, dispersive, magnetic, hysteretic, negative-index, etc. Any variation with time of the *total* electromagnetic momentum of a closed system results in a force exerted on the material media within the system in accordance with the generalized Lorentz law.

**1. Introduction**. In the classical electrodynamics of Maxwell and Lorentz, the so-called microscopic equations are treated as fundamental, while the roles of polarization density $\boldsymbol{P}(\boldsymbol{r},t)$ and magnetization density $\boldsymbol{M}(\boldsymbol{r},t)$ are considered secondary. The conventional wisdom is that, since, in the context of Maxwell's equations, $\boldsymbol{P}$ and $\boldsymbol{M}$ can be reduced to bound-charge and bound-current densities, there is no need to treat them as fundamentally different from free-charge and free-current densities, $\rho_{\text{free}}(\boldsymbol{r},t)$ and $\boldsymbol{J}_{\text{free}}(\boldsymbol{r},t)$. We believe, however, that $\boldsymbol{P}$ and $\boldsymbol{M}$ are not quite reducible to $\rho$ and $\boldsymbol{J}$ and that, therefore, they are of more fundamental significance than has heretofore been appreciated. The reasons for this belief do not lie within the Maxwell equations per se; rather it is the localized contributions of $\boldsymbol{P}$ and $\boldsymbol{M}$ to electromagnetic energy and force densities that differentiate polarization and magnetization from their "equivalent" bound-charge and bound-current densities.

In the bound-electric-charge picture, for example, one expects the force of the $E$-field on $\boldsymbol{P}(\boldsymbol{r},t)$ to be $\boldsymbol{F}(\boldsymbol{r},t) = -(\nabla \cdot \boldsymbol{P})\boldsymbol{E}$; experimental evidence, however, seems to suggest that the force density is $(\boldsymbol{P} \cdot \nabla)\boldsymbol{E}$ [1-5]. Even though the *total* force on a given object turns out to be the same in both formulations, the force *distributions* within the object and at its boundaries are generally quite different. The situation is even more intriguing in the case of magnetic media defined by their magnetization profile $\boldsymbol{M}(\boldsymbol{r},t)$. The well-known Lorentz law of force, $\boldsymbol{F} = q(\boldsymbol{E} + \boldsymbol{V} \times \boldsymbol{B})$, expressing the force exerted by the $\boldsymbol{E}$ and $\boldsymbol{B}$ fields on a particle of charge $q$ and velocity $\boldsymbol{V}$, is silent on the role of the magnetic moment of the particle, which generally originates from the spin and orbital angular momenta of the atomic constituents of matter. Treating magnetic dipoles as (bound) Amperian current loops would yield a force expression that is incompatible with the law of conservation of momentum [6], hence the occasional reports of "hidden momentum" in the literature over the past several decades [7-12].

However, if both $\boldsymbol{P}$ and $\boldsymbol{M}$ are treated as independent entities in their own rights, subject not only to Maxwell's macroscopic equations but also to an energy postulate and to a generalized form of the Lorentz law, one can show under very general circumstances that the total energy, momentum, and angular momentum of any closed system will be conserved. These conclusions in no way depend on $\boldsymbol{P}$ and $\boldsymbol{M}$ being linear functions of the local fields, on whether or not there is loss, anisotropy, nonlinearity, or hysteresis in the system, on whether the fields are propagating or evanescent, etc. In fact, $\boldsymbol{P}$ and $\boldsymbol{M}$ do not even have to be driven (i.e., excited) by the $\boldsymbol{E}$ and $\boldsymbol{H}$ fields: In principle, the space and time variations of $\boldsymbol{P}$ and $\boldsymbol{M}$ could altogether be dictated by non-electromagnetic forces and interactions such as mechanical force and quantum-mechanical exchange.

In this paper we demonstrate the conservation of energy and linear momentum under the most general circumstances conceivable in classical electrodynamics. The sole energy-related postulate that will be invoked is the definition of the Poynting vector $\boldsymbol{S}(\boldsymbol{r},t)$ as the rate of flow of electromagnetic energy per unit area per unit time. Also, we postulate that the electromagnetic momentum density is *always* given by $\boldsymbol{p}_{\text{EM}}(\boldsymbol{r},t) = \boldsymbol{S}(\boldsymbol{r},t)/c^2$, the so-called Abraham momentum density [13-15]. Armed with these postulates and with the following generalized form of the Lorenz law [1,5,6,14-18],

$$\boldsymbol{F}(\boldsymbol{r},t) = \rho_{\text{free}}\boldsymbol{E} + \boldsymbol{J}_{\text{free}} \times \mu_o\boldsymbol{H} + (\boldsymbol{P}\cdot\nabla)\boldsymbol{E} + (\boldsymbol{M}\cdot\nabla)\boldsymbol{H} + (\partial\boldsymbol{P}/\partial t)\times\mu_o\boldsymbol{H} - (\partial\boldsymbol{M}/\partial t)\times\varepsilon_o\boldsymbol{E}, \qquad (1)$$

we will prove that energy and linear momentum are universally conserved. A similar proof can be proffered for the conservation of angular momentum, but lack of space prevents us from presenting the proof in this paper.

As an example of application of the generalized Lorenz law given in Eq. (1), we present numerical simulation results pertaining to negative-index media, where $P$ and $M$ are related locally, isotropically, and linearly to the $E$ and $H$ fields, with the material's relative permittivity $\varepsilon(\omega)$ and permeability $\mu(\omega)$ both being negative-valued functions of frequency $\omega$. A slab of negative-index material subjected to radiation forces will be seen to behave in much the same way as a conventional (i.e., positive-index) medium would under similar circumstances. Our results, therefore, clearly contradict certain recently published theoretical predictions [19].

**2. Maxwell's equations in the Fourier domain**. In Maxwell's macroscopic equations, the sources of the electromagnetic fields $E(r,t)$ and $H(r,t)$ are the free-charge and free-current densities, $\rho_{\text{free}}(r,t)$ and $J_{\text{free}}(r,t)$, as well as the polarization and magnetization densities, $P(r,t)$ and $M(r,t)$. It will be assumed throughout the paper that the displacement field is $D(r,t) = \varepsilon_o E(r,t) + P(r,t)$, while the magnetic induction is $B(r,t) = \mu_o H(r,t) + M(r,t)$. In the MKSA system of units adopted here, the free-space permittivity $\varepsilon_o$ has units of farad/m, while the free-space permeability $\mu_o$ has units of henry/m. The units of the $E$- and $H$-fields are volt/m and ampere/m, respectively. Both $P$ and $D$ are in coulomb/m$^2$, while $B$ and $M$ have units of weber/m$^2$. The charge density distribution may be expanded in a Fourier integral as follows:

$$\rho_{\text{free}}(r,t) = (1/2\pi)^4 \int_{-\infty}^{\infty} \rho_{\text{free}}(k,\omega) \exp[i(k \cdot r - \omega t)] \, dk \, d\omega. \tag{2}$$

Similar Fourier transform relations can be written down for the remaining sources $J_{\text{free}}(r,t)$, $P(r,t)$, $M(r,t)$, as well as for the electromagnetic fields $E(r,t)$, $D(r,t)$, $H(r,t)$, and $B(r,t)$. Maxwell's macroscopic equations may then be transformed into the Fourier domain, and the scalar and vector potentials found as follows:

$$\psi(k,\omega) = \frac{\rho_{\text{free}}(k,\omega) - ik \cdot P(k,\omega)}{\varepsilon_o [k^2 - (\omega/c)^2]}; \tag{3a}$$

$$A(k,\omega) = \frac{\mu_o J_{\text{free}}(k,\omega) - i\mu_o \omega P(k,\omega) + ik \times M(k,\omega)}{k^2 - (\omega/c)^2}. \tag{3b}$$

The electromagnetic fields are then obtained from the above potentials as

$$E(k,\omega) = -ik\,\psi(k,\omega) + i\omega A(k,\omega)$$
$$= \{-i\mu_o c^2 [\rho_{\text{free}}(k,\omega) - ik \cdot P(k,\omega)]k + i\mu_o \omega [J_{\text{free}}(k,\omega) - i\omega P(k,\omega)] - \omega k \times M(k,\omega)\} / [k^2 - (\omega/c)^2]; \tag{4a}$$

$$D(k,\omega) = \varepsilon_o E(k,\omega) + P(k,\omega)$$
$$= \{-i[\rho_{\text{free}}(k,\omega) - ik \cdot P(k,\omega)]k + i(\omega/c^2)J_{\text{free}}(k,\omega) + k^2 P(k,\omega) - \varepsilon_o \omega k \times M(k,\omega)\} / [k^2 - (\omega/c)^2]; \tag{4b}$$

$$B(k,\omega) = ik \times A(k,\omega) = \{i\mu_o k \times J_{\text{free}}(k,\omega) + \mu_o \omega k \times P(k,\omega) + k^2 M(k,\omega) - [k \cdot M(k,\omega)]k\} / [k^2 - (\omega/c)^2]; \tag{4c}$$

$$H(k,\omega) = (1/\mu_o)[B(k,\omega) - M(k,\omega)] =$$
$$= \{ik \times J_{\text{free}}(k,\omega) + \omega k \times P(k,\omega) + \varepsilon_o \omega^2 M(k,\omega) - \varepsilon_o c^2 [k \cdot M(k,\omega)]k\} / [k^2 - (\omega/c)^2]. \tag{4d}$$

The above expressions for the $E$, $D$, $H$ and $B$ fields will now be used to derive formulas for the electromagnetic energy, force, and momentum of a closed system.

**3. Conservation of energy**. The Poynting vector $S(r,t) = E(r,t) \times H(r,t)$ and the local energy density $\mathcal{E}(r,t)$ of the electromagnetic field [5] satisfy the continuity equation $\nabla \cdot S(r,t) + \partial \mathcal{E}(r,t)/\partial t = 0$, where the time-rate-of-change of energy density is given by

$$\partial \mathcal{E}(r,t)/\partial t = E \cdot J_{\text{free}} + E \cdot \partial D/\partial t + H \cdot \partial B/\partial t. \tag{5}$$

The expression on the right-hand side of Eq. (5), when integrated over the entire space, turns out to be exactly equal to zero. In this way the total energy of any closed electromagnetic system satisfying Maxwell's macroscopic equations remains constant at all times. To prove this assertion, we substitute for the fields in Eq. (5) from Eqs. (4), then integrate over the entire space to find

$$(\partial/\partial t)\int_{-\infty}^{\infty}\mathcal{E}(\boldsymbol{r},t)\mathrm{d}\boldsymbol{r} = (1/2\pi)^8 \iint_{-\infty}^{\infty}[\boldsymbol{E}(\boldsymbol{k},\omega)\cdot\boldsymbol{J}_{\text{free}}(\boldsymbol{k}',\omega')-\mathrm{i}\omega'\boldsymbol{E}(\boldsymbol{k},\omega)\cdot\boldsymbol{D}(\boldsymbol{k}',\omega')-\mathrm{i}\omega\boldsymbol{H}(\boldsymbol{k}',\omega')\cdot\boldsymbol{B}(\boldsymbol{k},\omega)]$$

$$\times\left\{\int_{-\infty}^{\infty}\exp[\mathrm{i}(\boldsymbol{k}+\boldsymbol{k}')\cdot\boldsymbol{r}]\mathrm{d}\boldsymbol{r}\right\}\exp[-\mathrm{i}(\omega+\omega')t]\mathrm{d}\boldsymbol{k}\,\mathrm{d}\boldsymbol{k}'\mathrm{d}\omega\mathrm{d}\omega'. \quad (6)$$

With the help of the identity

$$\int_{-\infty}^{\infty}\exp[\mathrm{i}(\boldsymbol{k}+\boldsymbol{k}')\cdot\boldsymbol{r}]\mathrm{d}\boldsymbol{r} = (2\pi)^3\delta(\boldsymbol{k}+\boldsymbol{k}'), \quad (7)$$

where $\delta(\cdot)$ is Dirac's delta-function, Eq.(6) reduces to

$$(\partial/\partial t)\int_{-\infty}^{\infty}\mathcal{E}(\boldsymbol{r},t)\mathrm{d}\boldsymbol{r} =$$

$$(1/2\pi)^5\iint_{-\infty}^{\infty}\{\boldsymbol{E}(\boldsymbol{k},\omega)\cdot[\boldsymbol{J}_{\text{free}}(-\boldsymbol{k},\omega')-\mathrm{i}\omega'\boldsymbol{D}(-\boldsymbol{k},\omega')]-\mathrm{i}\omega\boldsymbol{H}(-\boldsymbol{k},\omega')\cdot\boldsymbol{B}(\boldsymbol{k},\omega)\}\exp[-\mathrm{i}(\omega+\omega')t]\mathrm{d}\boldsymbol{k}\,\mathrm{d}\omega\mathrm{d}\omega'. \quad (8)$$

Going forward, the calculation is straightforward but tedious. In the process we will need the charge continuity equation, $\nabla\cdot\boldsymbol{J}_{\text{free}}(\boldsymbol{r},t)+\partial\rho_{\text{free}}(\boldsymbol{r},t)/\partial t=0$, expressed in the Fourier domain as

$$\boldsymbol{k}\cdot\boldsymbol{J}_{\text{free}}(\boldsymbol{k},\omega) = \omega\rho_{\text{free}}(\boldsymbol{k},\omega). \quad (9)$$

Details of this calculation are presented in Appendix A. The final result is that the integrand of Eq.(8) is identically equal to zero and that, therefore, energy is precisely conserved. Note that, in deriving this result, no assumptions have been made about the nature of charge, current, polarization, and magnetization, above and beyond what has been implicit in Maxwell's macroscopic equations and also in Poynting's theorem, which equates the rate of flow of electromagnetic energy (per unit area per unit time) with $\boldsymbol{S}(\boldsymbol{r},t)$.

**4. Force and linear momentum**. Integrating the force density $\boldsymbol{F}(\boldsymbol{r},t)$ of Eq.(1) over the entire space, we find

$$\int_{-\infty}^{\infty}\boldsymbol{F}(\boldsymbol{r},t)\mathrm{d}\boldsymbol{r} = (1/2\pi)^8\iint_{-\infty}^{\infty}[\rho_{\text{free}}(\boldsymbol{k},\omega)\boldsymbol{E}(\boldsymbol{k}',\omega')+\boldsymbol{J}_{\text{free}}(\boldsymbol{k},\omega)\times\mu_{\text{o}}\boldsymbol{H}(\boldsymbol{k}',\omega')+\mathrm{i}[\boldsymbol{P}(\boldsymbol{k},\omega)\cdot\boldsymbol{k}']\boldsymbol{E}(\boldsymbol{k}',\omega')$$

$$-\mathrm{i}\omega\boldsymbol{P}(\boldsymbol{k},\omega)\times\mu_{\text{o}}\boldsymbol{H}(\boldsymbol{k}',\omega')+\mathrm{i}[\boldsymbol{M}(\boldsymbol{k},\omega)\cdot\boldsymbol{k}']\boldsymbol{H}(\boldsymbol{k}',\omega')+\mathrm{i}\omega\boldsymbol{M}(\boldsymbol{k},\omega)\times\varepsilon_{\text{o}}\boldsymbol{E}(\boldsymbol{k}',\omega')]$$

$$\times\left\{\int_{-\infty}^{\infty}\exp[\mathrm{i}(\boldsymbol{k}+\boldsymbol{k}')\cdot\boldsymbol{r}]\mathrm{d}\boldsymbol{r}\right\}\exp[-\mathrm{i}(\omega+\omega')t]\mathrm{d}\boldsymbol{k}\,\mathrm{d}\boldsymbol{k}'\mathrm{d}\omega\mathrm{d}\omega'. \quad (10)$$

Using the identity in Eq.(7), the above expression may be rewritten as follows:

$$\int_{-\infty}^{\infty}\boldsymbol{F}(\boldsymbol{r},t)\mathrm{d}\boldsymbol{r} = (1/2\pi)^5\iint_{-\infty}^{\infty}\{[\rho_{\text{free}}(\boldsymbol{k},\omega)-\mathrm{i}\boldsymbol{k}\cdot\boldsymbol{P}(\boldsymbol{k},\omega)]\boldsymbol{E}(-\boldsymbol{k},\omega')+\mathrm{i}\varepsilon_{\text{o}}\omega\boldsymbol{M}(\boldsymbol{k},\omega)\times\boldsymbol{E}(-\boldsymbol{k},\omega')$$

$$+\mu_{\text{o}}[\boldsymbol{J}_{\text{free}}(\boldsymbol{k},\omega)-\mathrm{i}\omega\boldsymbol{P}(\boldsymbol{k},\omega)]\times\boldsymbol{H}(-\boldsymbol{k},\omega')-\mathrm{i}[\boldsymbol{k}\cdot\boldsymbol{M}(\boldsymbol{k},\omega)]\boldsymbol{H}(-\boldsymbol{k},\omega')\}\exp[-\mathrm{i}(\omega+\omega')t]\mathrm{d}\boldsymbol{k}\,\mathrm{d}\omega\mathrm{d}\omega'. \quad (11)$$

Next we evaluate the time-rate-of-change of the total electromagnetic momentum of the system.

$$(\partial/\partial t)\int_{-\infty}^{\infty}(1/c^2)\boldsymbol{S}(\boldsymbol{r},t)\mathrm{d}\boldsymbol{r} = (1/2\pi)^5\iint_{-\infty}^{\infty}[-\mathrm{i}(\omega+\omega')/c^2]\boldsymbol{E}(\boldsymbol{k},\omega)\times\boldsymbol{H}(-\boldsymbol{k},\omega')\exp[-\mathrm{i}(\omega+\omega')t]\mathrm{d}\boldsymbol{k}\,\mathrm{d}\omega\mathrm{d}\omega'. \quad (12)$$

Substituting for the $\boldsymbol{E}$ and $\boldsymbol{H}$ fields from Eqs.(4), and using the charge continuity condition of Eq.(9), one can readily show that the total force of Eq.(11) is exactly equal and opposite in sign to the time-derivative of the total momentum given by Eq.(12); for a detailed proof see Appendix B. Consequently,

$$\int_{-\infty}^{\infty}\boldsymbol{F}(\boldsymbol{r},t)\mathrm{d}\boldsymbol{r} + (\partial/\partial t)\int_{-\infty}^{\infty}\boldsymbol{p}_{\text{EM}}(\boldsymbol{r},t)\mathrm{d}\boldsymbol{r} = 0. \quad (13)$$

In words, at any instant of time *t*, the *total* force exerted on the material media within any *closed* system is the negative of the time-rate-of-change of the *total* electromagnetic momentum of the system at that instant. This, of course, is nothing more nor less than the statement of conservation of linear momentum in electromagnetic systems. Note once again that in deriving Eq.(13) no assumptions were made about the nature of charge, current, polarization, and magnetization, above and beyond what has been implicit in Maxwell's macroscopic equations and in the generalized Lorentz law of force as given by Eq.(1). The only other assumption that was needed in arriving at Eq.(13) is the postulate that electromagnetic momentum density, $\boldsymbol{p}_{\text{EM}}(\boldsymbol{r},t)$, is *always* given by $\boldsymbol{S}(\boldsymbol{r},t)/c^2$.

**5. Numerical simulations**. Using the finite difference time domain (FDTD) algorithm, we solved Maxwell's macroscopic equations in specific situations involving homogeneous, isotropic, linear, dispersive media in which $\boldsymbol{P}(\boldsymbol{r},\omega) = \varepsilon_{\text{o}}[\varepsilon(\omega)-1]\boldsymbol{E}(\boldsymbol{r},\omega)$ and $\boldsymbol{M}(\boldsymbol{r},\omega) = \mu_{\text{o}}[\mu(\omega)-1]\boldsymbol{H}(\boldsymbol{r},\omega)$. We then computed the distribution of the Lorentz force in accordance with Eq. (1), and that of the electromagnetic momentum using $p_{EM}(r,t)=E(r,t)\times H(r,t)/c^2$. It was confirmed that Eq. (13) is satisfied in each and every case.

As an example, Fig. 1(a) shows a convergent beam of light (vacuum wavelength $\lambda_o=400$ nm, sourced at $z=3\,\mu$m) propagating in the free space along the negative $z$-axis. The axis of the beam is inclined at about 30° toward the negative $y$-axis. In this 2D simulation, the beam, being uniform along the $x$-axis, is linearly polarized in the $yz$-plane. The field components, therefore, are ($E_y$, $E_z$, $H_x$), with the instantaneous profile of $H_x$ shown in Fig. 1(a). Later we placed in the path of the beam a semi-infinite slab of a hypothetical negative-index material, filling the region $z \leq 2.5\,\mu$m. The complex-valued permittivity $\varepsilon(\omega)$ and permeability $\mu(\omega)$ of the slab material over the wavelength range $\lambda_o = 355 - 425$ nm are displayed in Fig. 1(b). The real parts of $\varepsilon$ and $\mu$ (solid curves) are negative over this entire range of wavelengths, while their (small) imaginary parts, shown as dashed curves, are responsible for the absorption of the light as the beam propagates inside the medium.

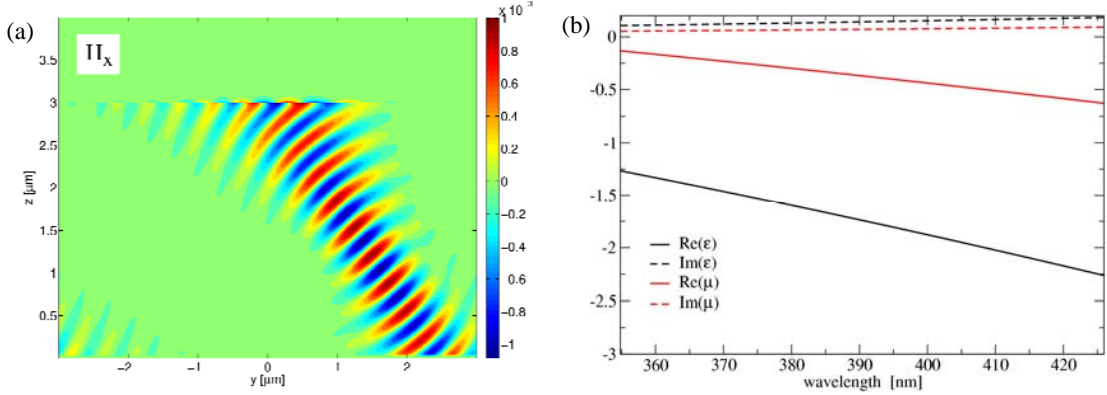

**Fig. 1**. (a) A convergent beam of light ($\lambda_o=400$ nm, sourced at $z=3\,\mu$m) propagates in free space along the negative $z$-axis, with ~30° tilt away from $z$. In this 2D simulation, the beam, whose profile is uniform along the $x$-axis, is linearly polarized in the $yz$-plane; also periodic boundary conditions are applied along the $y$-axis. Of the three field components ($E_y$, $E_z$, $H_x$) only the instantaneous profile of $H_x$ is shown. (b) Plots of the complex-valued dielectric permittivity $\varepsilon(\omega)$ and magnetic permeability $\mu(\omega)$ of a hypothetical negative-index material, a slab of which is to be placed in the path of the light beam depicted in (a). The wavelength dependences of $\varepsilon$ and $\mu$ are shown over the wavelength range $\lambda_o = 355 - 425$ nm.

Figure 2(a) shows a semi-infinite slab of the negative-index material placed in the path of the incident beam of Fig. 1(a); the top of the slab is located at $z=2.5\,\mu$m. Since the incident beam is sourced at $z=3\,\mu$m, interference between the incident and reflected beams occurs only in the narrow region below the source plane and above the slab. Above this narrow band, in the region $z \geq 3\,\mu$m, the reflected beam, whose field components are shown in Figs. 2(b-d), can be seen propagating away from the slab. Inside the negative-index medium and just below its surface, where $z \leq 2.5\,\mu$m, the fraction of the beam that enters the slab bends away from the surface normal on the side of the incident beam, a characteristic of refraction into a negative-index medium. Once inside the slab, the imaginary parts of $\varepsilon$ and $\mu$ ensure that the beam is quickly attenuated along the direction of propagation.

Instantaneous profiles of the force components $F_y$ and $F_z$ inside and at the entrance facet of the negative-index medium are shown in Fig. 3, while the time-averaged force profiles $<F_y>$ and $<F_z>$ appear in Fig. 4. Note that a substantial $z$-component of force exists at the entrance facet of the slab. The total force $<\int F_z \mathrm{d}r>$ exerted on the slab in the $z$-direction is negative, meaning that the incident light pushes the medium downward. The magnitude of this force is in precise numerical agreement with the net rate of momentum transfer to the slab by the incident and reflected beams, both of which reside in the free space and, therefore, carry pure electromagnetic momentum.

The time-averaged force $<F_y>$ in the horizontal direction, depicted in Fig. 4(a), is positive on the right-hand side and negative on the left-hand side of the beam, meaning that the force tries to expand the medium horizontally. The net horizontal force experienced by the slab is positive, which indicates that the slab is being pushed to the right. Once again, this is consistent with momentum conservation, as the horizontal momentum of the incident beam is greater than that of the reflected beam – both beams, of course, reside in the free space and carry electromagnetic momentum only. The difference between the rates of arrival and departure of electromagnetic momentum at the surface of the slab (due to the $y$-directed components of incident and reflected light) is found to be numerically identical to the total horizontal force $<\int F_y \mathrm{d}r>$ experienced by the slab.

**Acknowledgements.** This work has been supported by the Air Force Office of Scientific Research (AFOSR) under contract number FA 9550−04−1−0213.

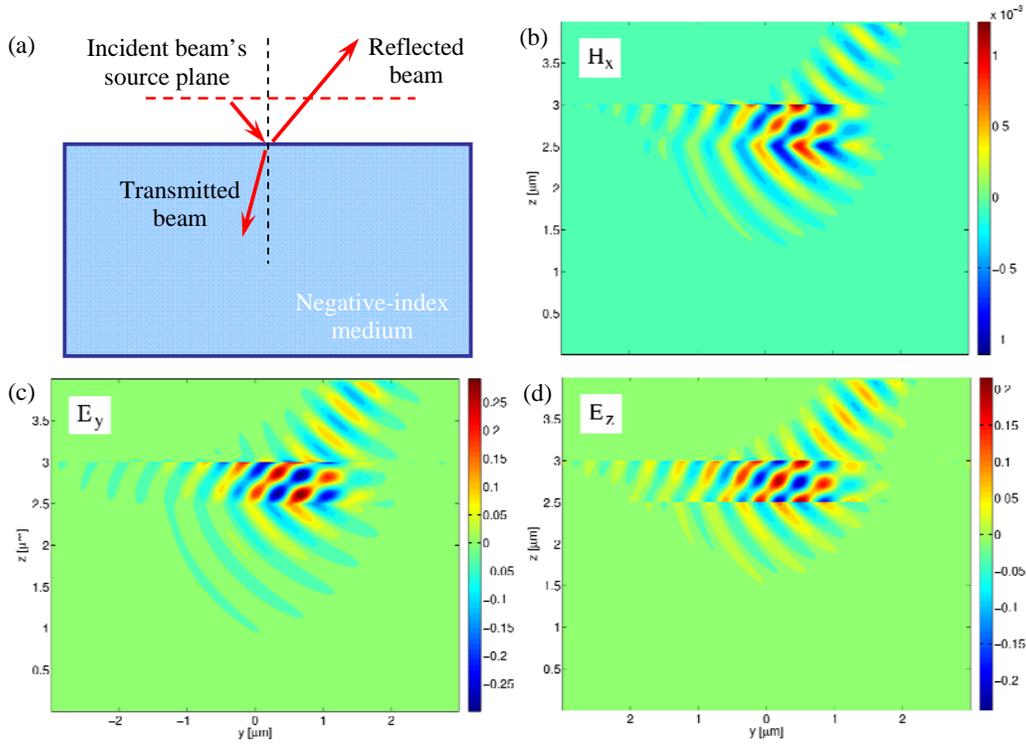

**Fig. 2**. A semi-infinite slab of a negative-index material is placed in the path of the beam depicted in Fig. 1(a). The slab surface is at $z = 2.5\,\mu$m. Since the incident beam is sourced at $z = 3.0\,\mu$m, interference between the incident and reflected beams occurs only in the band $2.5 \leq z \leq 3\,\mu$m. Above this band, where $z \geq 3\,\mu$m, only the reflected beam is visible. The beam transmitted into the slab (and rather quickly absorbed) appears in the region $z \leq 2.5\,\mu$m.

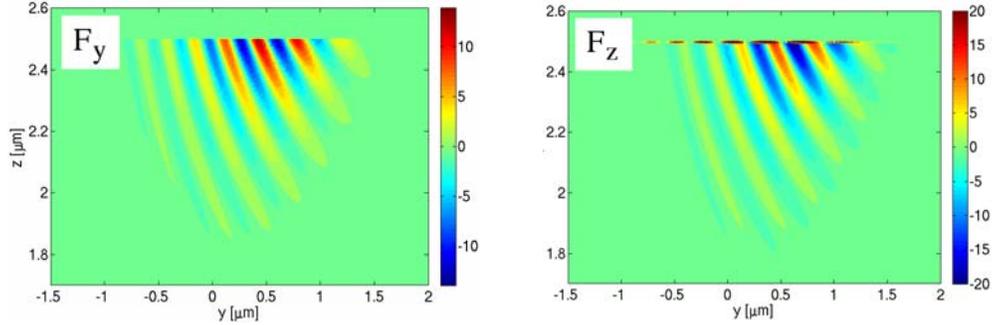

**Fig. 3**. Instantaneous profiles of the force components $F_y$ and $F_z$ inside and at the entrance facet of the negative-index medium depicted in Fig. 2.

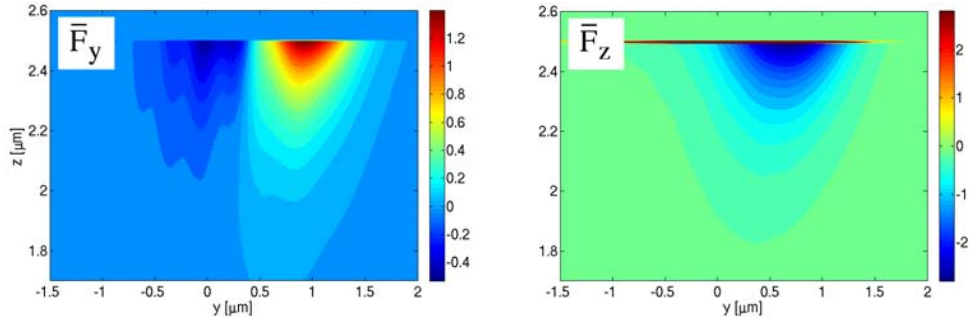

**Fig. 4**. Profiles of time-averaged force components $\langle F_y \rangle$ and $\langle F_z \rangle$ experienced by the negative-index medium depicted in Fig. 2. Note that a substantial $z$-component of force exists at the entrance facet of the slab. The overall force of the light beam (integrated over the bulk as well as the surface of the slab) tends to push the slab to the right and downward, in agreement with the time-rate-of-change of the momenta delivered to the slab by the incident and reflected beams.

# Appendix A

We expand the integrand of Eq.(8) by substituting for the fields from Eqs.(4), then combining and rearranging the various terms until, in the final expression, each remaining term has an equal but opposite counterpart.

$$\boldsymbol{E}(\boldsymbol{k},\omega) \cdot [\boldsymbol{J}_{\text{free}}(-\boldsymbol{k},\omega') - i\omega' \boldsymbol{D}(-\boldsymbol{k},\omega')] - i\omega \boldsymbol{H}(-\boldsymbol{k},\omega') \cdot \boldsymbol{B}(\boldsymbol{k},\omega) = \big\{ \{-\mu_o c^2 [\boldsymbol{k}\cdot\boldsymbol{P}(\boldsymbol{k},\omega) + i\rho_{\text{free}}(\boldsymbol{k},\omega)]\boldsymbol{k}$$
$$+ i\mu_o \omega \boldsymbol{J}_{\text{free}}(\boldsymbol{k},\omega) + \mu_o \omega^2 \boldsymbol{P}(\boldsymbol{k},\omega) - \omega \boldsymbol{k}\times\boldsymbol{M}(\boldsymbol{k},\omega)\} \cdot \{\omega'[i\boldsymbol{k}\cdot\boldsymbol{P}(-\boldsymbol{k},\omega') + \rho_{\text{free}}(-\boldsymbol{k},\omega')]\boldsymbol{k} + k^2 \boldsymbol{J}_{\text{free}}(-\boldsymbol{k},\omega')$$
$$- i\omega' k^2 \boldsymbol{P}(-\boldsymbol{k},\omega') - i\varepsilon_o \omega'^2 \boldsymbol{k}\times\boldsymbol{M}(-\boldsymbol{k},\omega')\} + \{-i\boldsymbol{k}\times\boldsymbol{J}_{\text{free}}(-\boldsymbol{k},\omega') - \omega'\boldsymbol{k}\times\boldsymbol{P}(-\boldsymbol{k},\omega') - \varepsilon_o c^2 [\boldsymbol{k}\cdot\boldsymbol{M}(-\boldsymbol{k},\omega')]\boldsymbol{k}$$
$$+ \varepsilon_o \omega'^2 \boldsymbol{M}(-\boldsymbol{k},\omega')\} \cdot \{\mu_o \omega \boldsymbol{k}\times\boldsymbol{J}_{\text{free}}(\boldsymbol{k},\omega) - i\mu_o \omega^2 \boldsymbol{k}\times\boldsymbol{P}(\boldsymbol{k},\omega) + i\omega[\boldsymbol{k}\cdot\boldsymbol{M}(\boldsymbol{k},\omega)]\boldsymbol{k} - i\omega k^2 \boldsymbol{M}(\boldsymbol{k},\omega)\} \big\} /$$
$$\{[k^2 - (\omega/c)^2][k^2 - (\omega'/c)^2]\}$$

$$= \{-\mu_o c^2 \omega' k^2 [\boldsymbol{k}\cdot\boldsymbol{P}(\boldsymbol{k},\omega) + i\rho_{\text{free}}(\boldsymbol{k},\omega)][i\boldsymbol{k}\cdot\boldsymbol{P}(-\boldsymbol{k},\omega') + \rho_{\text{free}}(-\boldsymbol{k},\omega')]$$
$$- \mu_o c^2 k^2 [\boldsymbol{k}\cdot\boldsymbol{P}(\boldsymbol{k},\omega) + i\rho_{\text{free}}(\boldsymbol{k},\omega)]\boldsymbol{k}\cdot\boldsymbol{J}_{\text{free}}(-\boldsymbol{k},\omega') + i\mu_o c^2 \omega' k^2 [\boldsymbol{k}\cdot\boldsymbol{P}(\boldsymbol{k},\omega) + i\rho_{\text{free}}(\boldsymbol{k},\omega)]\boldsymbol{k}\cdot\boldsymbol{P}(-\boldsymbol{k},\omega')$$
$$+ i\mu_o \omega \omega' [i\boldsymbol{k}\cdot\boldsymbol{P}(-\boldsymbol{k},\omega') + \rho_{\text{free}}(-\boldsymbol{k},\omega')]\boldsymbol{k}\cdot\boldsymbol{J}_{\text{free}}(\boldsymbol{k},\omega) + i\mu_o \omega k^2 \boldsymbol{J}_{\text{free}}(\boldsymbol{k},\omega) \cdot \boldsymbol{J}_{\text{free}}(-\boldsymbol{k},\omega')$$
$$+ \mu_o \omega \omega' k^2 \boldsymbol{J}_{\text{free}}(\boldsymbol{k},\omega)\cdot\boldsymbol{P}(-\boldsymbol{k},\omega') + \omega(\omega'/c)^2 [\boldsymbol{k}\times\boldsymbol{M}(-\boldsymbol{k},\omega')]\cdot\boldsymbol{J}_{\text{free}}(\boldsymbol{k},\omega)$$
$$+ \mu_o \omega^2 \omega' [i\boldsymbol{k}\cdot\boldsymbol{P}(-\boldsymbol{k},\omega') + \rho_{\text{free}}(-\boldsymbol{k},\omega')]\boldsymbol{k}\cdot\boldsymbol{P}(\boldsymbol{k},\omega) + \mu_o \omega^2 k^2 \boldsymbol{P}(\boldsymbol{k},\omega)\cdot\boldsymbol{J}_{\text{free}}(-\boldsymbol{k},\omega')$$
$$- i\mu_o \omega^2 \omega' k^2 \boldsymbol{P}(\boldsymbol{k},\omega)\cdot\boldsymbol{P}(-\boldsymbol{k},\omega') - i(\omega/c)^2 \omega'^2 [\boldsymbol{k}\times\boldsymbol{M}(-\boldsymbol{k},\omega')]\cdot\boldsymbol{P}(\boldsymbol{k},\omega) - \omega k^2 [\boldsymbol{k}\times\boldsymbol{M}(\boldsymbol{k},\omega)]\cdot\boldsymbol{J}_{\text{free}}(-\boldsymbol{k},\omega')$$
$$+ i\omega \omega' k^2 [\boldsymbol{k}\times\boldsymbol{M}(\boldsymbol{k},\omega)]\cdot\boldsymbol{P}(-\boldsymbol{k},\omega') + i\varepsilon_o \omega \omega'^2 [\boldsymbol{k}\times\boldsymbol{M}(-\boldsymbol{k},\omega')]\cdot[\boldsymbol{k}\times\boldsymbol{M}(\boldsymbol{k},\omega)]$$
$$- i\mu_o \omega [\boldsymbol{k}\times\boldsymbol{J}_{\text{free}}(\boldsymbol{k},\omega)]\cdot[\boldsymbol{k}\times\boldsymbol{J}_{\text{free}}(-\boldsymbol{k},\omega')] - \mu_o \omega^2 [\boldsymbol{k}\times\boldsymbol{P}(\boldsymbol{k},\omega)]\cdot[\boldsymbol{k}\times\boldsymbol{J}_{\text{free}}(-\boldsymbol{k},\omega')]$$
$$- \omega k^2 [\boldsymbol{k}\times\boldsymbol{J}_{\text{free}}(-\boldsymbol{k},\omega')]\cdot\boldsymbol{M}(\boldsymbol{k},\omega) - \mu_o \omega \omega' [\boldsymbol{k}\times\boldsymbol{J}_{\text{free}}(\boldsymbol{k},\omega)]\cdot[\boldsymbol{k}\times\boldsymbol{P}(-\boldsymbol{k},\omega')]$$
$$+ i\mu_o \omega^2 \omega' [\boldsymbol{k}\times\boldsymbol{P}(\boldsymbol{k},\omega)]\cdot[\boldsymbol{k}\times\boldsymbol{P}(-\boldsymbol{k},\omega')] + i\omega \omega' k^2 [\boldsymbol{k}\times\boldsymbol{P}(-\boldsymbol{k},\omega')]\cdot\boldsymbol{M}(\boldsymbol{k},\omega)$$
$$- i\varepsilon_o c^2 \omega k^2 [\boldsymbol{k}\cdot\boldsymbol{M}(\boldsymbol{k},\omega)][\boldsymbol{k}\cdot\boldsymbol{M}(-\boldsymbol{k},\omega')] + i\varepsilon_o c^2 \omega k^2 [\boldsymbol{k}\cdot\boldsymbol{M}(\boldsymbol{k},\omega)][\boldsymbol{k}\cdot\boldsymbol{M}(-\boldsymbol{k},\omega')]$$
$$+ \omega(\omega'/c)^2 [\boldsymbol{k}\times\boldsymbol{J}_{\text{free}}(\boldsymbol{k},\omega)]\cdot\boldsymbol{M}(-\boldsymbol{k},\omega') - i(\omega \omega'/c)^2 [\boldsymbol{k}\times\boldsymbol{P}(\boldsymbol{k},\omega)]\cdot\boldsymbol{M}(-\boldsymbol{k},\omega')$$
$$+ i\varepsilon_o \omega \omega'^2 [\boldsymbol{k}\cdot\boldsymbol{M}(\boldsymbol{k},\omega)][\boldsymbol{k}\cdot\boldsymbol{M}(-\boldsymbol{k},\omega')] - i\varepsilon_o \omega \omega'^2 k^2 \boldsymbol{M}(\boldsymbol{k},\omega)\cdot\boldsymbol{M}(-\boldsymbol{k},\omega')\} / \{[k^2-(\omega/c)^2][k^2-(\omega'/c)^2]\}$$

$$= \{+i\mu_o \omega k^2 \boldsymbol{J}_{\text{free}}(\boldsymbol{k},\omega)\cdot\boldsymbol{J}_{\text{free}}(-\boldsymbol{k},\omega') - i\mu_o \omega [\boldsymbol{k}\cdot\boldsymbol{J}_{\text{free}}(\boldsymbol{k},\omega)][\boldsymbol{k}\cdot\boldsymbol{J}_{\text{free}}(-\boldsymbol{k},\omega')]$$
$$- i\mu_o \omega^2 \omega' k^2 \boldsymbol{P}(\boldsymbol{k},\omega)\cdot\boldsymbol{P}(-\boldsymbol{k},\omega') + i\mu_o \omega^2 \omega' [\boldsymbol{k}\cdot\boldsymbol{P}(\boldsymbol{k},\omega)][\boldsymbol{k}\cdot\boldsymbol{P}(-\boldsymbol{k},\omega')]$$
$$+ i\varepsilon_o \omega \omega'^2 k^2 \boldsymbol{M}(\boldsymbol{k},\omega)\cdot\boldsymbol{M}(-\boldsymbol{k},\omega') - i\varepsilon_o \omega \omega'^2 [\boldsymbol{k}\cdot\boldsymbol{M}(\boldsymbol{k},\omega)][\boldsymbol{k}\cdot\boldsymbol{M}(-\boldsymbol{k},\omega')]$$
$$+ \mu_o \omega \omega' k^2 \boldsymbol{J}_{\text{free}}(\boldsymbol{k},\omega)\cdot\boldsymbol{P}(-\boldsymbol{k},\omega') - \mu_o \omega \omega' [\boldsymbol{k}\cdot\boldsymbol{J}_{\text{free}}(\boldsymbol{k},\omega)][\boldsymbol{k}\cdot\boldsymbol{P}(-\boldsymbol{k},\omega')]$$
$$+ \mu_o \omega^2 k^2 \boldsymbol{P}(\boldsymbol{k},\omega)\cdot\boldsymbol{J}_{\text{free}}(-\boldsymbol{k},\omega') - \mu_o \omega^2 [\boldsymbol{k}\cdot\boldsymbol{P}(\boldsymbol{k},\omega)][\boldsymbol{k}\cdot\boldsymbol{J}_{\text{free}}(-\boldsymbol{k},\omega')]$$
$$- \omega(\omega'/c)^2 [\boldsymbol{k}\times\boldsymbol{J}_{\text{free}}(\boldsymbol{k},\omega)]\cdot\boldsymbol{M}(-\boldsymbol{k},\omega') + \omega k^2 [\boldsymbol{k}\times\boldsymbol{J}_{\text{free}}(-\boldsymbol{k},\omega')]\cdot\boldsymbol{M}(\boldsymbol{k},\omega)$$
$$+ i(\omega \omega'/c)^2 [\boldsymbol{k}\times\boldsymbol{P}(\boldsymbol{k},\omega)]\cdot\boldsymbol{M}(-\boldsymbol{k},\omega') - i\omega \omega' k^2 [\boldsymbol{k}\times\boldsymbol{P}(-\boldsymbol{k},\omega')]\cdot\boldsymbol{M}(\boldsymbol{k},\omega)$$
$$- i\mu_o \omega k^2 \boldsymbol{J}_{\text{free}}(\boldsymbol{k},\omega)\cdot\boldsymbol{J}_{\text{free}}(-\boldsymbol{k},\omega') + i\mu_o \omega [\boldsymbol{k}\cdot\boldsymbol{J}_{\text{free}}(\boldsymbol{k},\omega)][\boldsymbol{k}\cdot\boldsymbol{J}_{\text{free}}(-\boldsymbol{k},\omega')]$$
$$- \mu_o \omega^2 k^2 \boldsymbol{P}(\boldsymbol{k},\omega)\cdot\boldsymbol{J}_{\text{free}}(-\boldsymbol{k},\omega') + \mu_o \omega^2 [\boldsymbol{k}\cdot\boldsymbol{P}(\boldsymbol{k},\omega)][\boldsymbol{k}\cdot\boldsymbol{J}_{\text{free}}(-\boldsymbol{k},\omega')]$$
$$- \mu_o \omega \omega' k^2 \boldsymbol{J}_{\text{free}}(\boldsymbol{k},\omega)\cdot\boldsymbol{P}(-\boldsymbol{k},\omega') + \mu_o \omega \omega' [\boldsymbol{k}\cdot\boldsymbol{J}_{\text{free}}(\boldsymbol{k},\omega)][\boldsymbol{k}\cdot\boldsymbol{P}(-\boldsymbol{k},\omega')]$$
$$+ i\mu_o \omega^2 \omega' k^2 \boldsymbol{P}(\boldsymbol{k},\omega)\cdot\boldsymbol{P}(-\boldsymbol{k},\omega') - i\mu_o \omega^2 \omega' [\boldsymbol{k}\cdot\boldsymbol{P}(\boldsymbol{k},\omega)][\boldsymbol{k}\cdot\boldsymbol{P}(-\boldsymbol{k},\omega')]$$
$$+ i\omega \omega' k^2 [\boldsymbol{k}\times\boldsymbol{P}(-\boldsymbol{k},\omega')]\cdot\boldsymbol{M}(\boldsymbol{k},\omega) - i(\omega \omega'/c)^2 [\boldsymbol{k}\times\boldsymbol{P}(\boldsymbol{k},\omega)]\cdot\boldsymbol{M}(-\boldsymbol{k},\omega')$$
$$+ \omega(\omega'/c)^2 [\boldsymbol{k}\times\boldsymbol{J}_{\text{free}}(\boldsymbol{k},\omega)]\cdot\boldsymbol{M}(-\boldsymbol{k},\omega') - \omega k^2 [\boldsymbol{k}\times\boldsymbol{J}_{\text{free}}(-\boldsymbol{k},\omega')]\cdot\boldsymbol{M}(\boldsymbol{k},\omega)$$
$$+ i\varepsilon_o \omega \omega'^2 [\boldsymbol{k}\cdot\boldsymbol{M}(\boldsymbol{k},\omega)][\boldsymbol{k}\cdot\boldsymbol{M}(-\boldsymbol{k},\omega')] - i\varepsilon_o \omega \omega'^2 k^2 \boldsymbol{M}(\boldsymbol{k},\omega)\cdot\boldsymbol{M}(-\boldsymbol{k},\omega')\} / \{[k^2-(\omega/c)^2][k^2-(\omega'/c)^2]\}.$$

In the final expression, all the various terms cancel out and the net result is precisely equal to zero.

# Appendix B

Upon substitution for *E* and *H*-fields from Eqs.(4), Eq.(11) may be written

$$\int_{-\infty}^{\infty} F(r,t)\,dr = (1/2\pi)^5 \iiint_{-\infty}^{\infty} [k^2-(\omega'/c)^2]^{-1}\{i\mu_o c^2\{[\rho_{\text{free}}(k,\omega)-ik\cdot P(k,\omega)][\rho_{\text{free}}(-k,\omega')+ik\cdot P(-k,\omega')]\}k$$

$$-i\mu_o\{[J_{\text{free}}(k,\omega)-i\omega P(k,\omega)]\cdot[J_{\text{free}}(-k,\omega')-i\omega' P(-k,\omega')]\}k$$

$$+i\mu_o(\omega+\omega')\{[\rho_{\text{free}}(k,\omega)-ik\cdot P(k,\omega)][J_{\text{free}}(-k,\omega')-i\omega' P(-k,\omega')]\}$$

$$-i\varepsilon_o\omega'(\omega+\omega')[k\cdot M(k,\omega)]M(-k,\omega')$$

$$+\omega[\rho_{\text{free}}(-k,\omega')+ik\cdot P(-k,\omega')][k\times M(k,\omega)]+\omega'[\rho_{\text{free}}(k,\omega)-ik\cdot P(k,\omega)][k\times M(-k,\omega')]$$

$$+(\omega'/c)^2[J_{\text{free}}(k,\omega)-i\omega P(k,\omega)]\times M(-k,\omega')+(\omega\omega'/c^2)[J_{\text{free}}(-k,\omega')-i\omega' P(-k,\omega')]\times M(k,\omega)$$

$$+[k\cdot M(-k,\omega')]\{k\times[J_{\text{free}}(k,\omega)-i\omega P(k,\omega)]\}-[k\cdot M(k,\omega)]\{k\times[J_{\text{free}}(-k,\omega')-i\omega' P(-k,\omega')]\}$$

$$+i\varepsilon_o c^2[k\cdot M(k,\omega)][k\cdot M(-k,\omega')]k+i\varepsilon_o\omega\omega'[M(k,\omega)\cdot M(-k,\omega')]k\}$$

$$\times\exp[-i(\omega+\omega')t]\,dk\,d\omega\,d\omega'. \quad (B1)$$

Some of the terms in the above expression can be further simplified if the integral is taken once over $(k,\omega,\omega')$, a second time over $(-k,\omega',\omega)$, then the results added together and normalized by 2. That the two integrals should be identical is the obvious consequence of a simple change of variables. We will have

$$\int_{-\infty}^{\infty} F(r,t)\,dr = (1/2\pi)^5 \iiint_{-\infty}^{\infty} \{-\tfrac{1}{2}i\mu_o(\omega-\omega')\{[\rho_{\text{free}}(k,\omega)-ik\cdot P(k,\omega)][\rho_{\text{free}}(-k,\omega')+ik\cdot P(-k,\omega')]\}k$$

$$+\tfrac{1}{2}i\mu_o[(\omega-\omega')/c^2]\{[J_{\text{free}}(k,\omega)-i\omega P(k,\omega)]\cdot[J_{\text{free}}(-k,\omega')-i\omega' P(-k,\omega')]\}k$$

$$+i\mu_o[k^2-(\omega/c)^2]\{[\rho_{\text{free}}(k,\omega)-ik\cdot P(k,\omega)][J_{\text{free}}(-k,\omega')-i\omega' P(-k,\omega')]\}$$

$$-i\varepsilon_o\omega'[k^2-(\omega/c)^2][k\cdot M(k,\omega)]M(-k,\omega')$$

$$-[\omega'(\omega-\omega')/c^2][\rho_{\text{free}}(k,\omega)-ik\cdot P(k,\omega)][k\times M(-k,\omega')]$$

$$+(\omega'/c^2)[k^2-(\omega\omega'/c^2)][J_{\text{free}}(k,\omega)-i\omega P(k,\omega)]\times M(-k,\omega')$$

$$-[(\omega-\omega')/c^2][k\cdot M(-k,\omega')]\{k\times[J_{\text{free}}(k,\omega)-i\omega P(k,\omega)]\}$$

$$-\tfrac{1}{2}i\varepsilon_o(\omega-\omega')[k\cdot M(k,\omega)][k\cdot M(-k,\omega')]k$$

$$-\tfrac{1}{2}i\varepsilon_o[\omega\omega'(\omega-\omega')/c^2][M(k,\omega)\cdot M(-k,\omega')]k\}(\omega+\omega')$$

$$\times\exp[-i(\omega+\omega')t]/\{[k^2-(\omega/c)^2][k^2-(\omega'/c)^2]\}\,dk\,d\omega\,d\omega'. \quad (B2)$$

Next we evaluate Eq.(12) by substituting for the *E* and *H* fields from Eqs.(4).

$$(\partial/\partial t)\int_{-\infty}^{\infty}(1/c^2)S(r,t)\,dr$$

$$=(1/2\pi)^5\iiint_{-\infty}^{\infty}\{-i\mu_o[k^2-(\omega/c)^2]\rho_{\text{free}}(k,\omega)J_{\text{free}}(-k,\omega')-\mu_o[k^2-(\omega/c)^2]\omega'\rho_{\text{free}}(k,\omega)P(-k,\omega')$$

$$-\mu_o[k^2-(\omega/c)^2][k\cdot P(k,\omega)]J_{\text{free}}(-k,\omega')+i\mu_o[k^2-(\omega/c)^2]\omega'[k\cdot P(k,\omega)]P(-k,\omega')$$

$$-i\varepsilon_o\omega[k^2-(\omega'/c)^2][k\cdot M(-k,\omega')]M(k,\omega)$$

$$-i\mu_o\omega'[\rho_{\text{free}}(k,\omega)-ik\cdot P(k,\omega)][\rho_{\text{free}}(-k,\omega')+ik\cdot P(-k,\omega')]k$$

$$-i\mu_o(\omega/c^2)\{[J_{\text{free}}(k,\omega)-i\omega P(k,\omega)]\cdot[J_{\text{free}}(-k,\omega')-i\omega' P(-k,\omega')]\}k$$

$$-i\varepsilon_o\omega(\omega'/c)^2[M(k,\omega)\cdot M(-k,\omega')]k+i\varepsilon_o\omega[k\cdot M(k,\omega)][k\cdot M(-k,\omega')]k$$

$$+(\omega/c^2)(\omega'/c)^2[J_{\text{free}}(k,\omega)-i\omega P(k,\omega)]\times M(-k,\omega')$$

$$-(\omega'/c)^2[\rho_{\text{free}}(k,\omega)-ik\cdot P(k,\omega)][k\times M(-k,\omega')]$$

$$+ (\omega/c^2)[\mathbf{k}\cdot\mathbf{M}(-\mathbf{k},\omega')]\{\mathbf{k}\times[\mathbf{J}_{\text{free}}(\mathbf{k},\omega)-\mathrm{i}\omega\mathbf{P}(\mathbf{k},\omega)]\}$$
$$+ (\omega/c^2)\{[\mathbf{k}\times\mathbf{M}(\mathbf{k},\omega)]\cdot[\mathbf{J}_{\text{free}}(-\mathbf{k},\omega')-\mathrm{i}\omega'\mathbf{P}(-\mathbf{k},\omega')]\}\mathbf{k}\,\}(\omega+\omega')$$
$$\times \exp[-\mathrm{i}(\omega+\omega')t]/\{[k^2-(\omega/c)^2][k^2-(\omega'/c)^2]\}\,\mathrm{d}\mathbf{k}\,\mathrm{d}\omega\,\mathrm{d}\omega'. \tag{B3}$$

Once again we combine the terms using the equality of integrals taken over $(\mathbf{k},\omega,\omega')$ and $(-\mathbf{k},\omega',\omega)$ to obtain:

$$= (1/2\pi)^5 \iiint_{-\infty}^{\infty} \{-\mathrm{i}\mu_0[k^2-(\omega/c)^2][\rho_{\text{free}}(\mathbf{k},\omega)-\mathrm{i}\mathbf{k}\cdot\mathbf{P}(\mathbf{k},\omega)][\mathbf{J}_{\text{free}}(-\mathbf{k},\omega')-\mathrm{i}\omega'\mathbf{P}(-\mathbf{k},\omega')]$$
$$+ \tfrac{1}{2}\mathrm{i}\mu_0(\omega-\omega')[\rho_{\text{free}}(\mathbf{k},\omega)-\mathrm{i}\mathbf{k}\cdot\mathbf{P}(\mathbf{k},\omega)][\rho_{\text{free}}(-\mathbf{k},\omega')+\mathrm{i}\mathbf{k}\cdot\mathbf{P}(-\mathbf{k},\omega')]\mathbf{k}$$
$$- \tfrac{1}{2}\mathrm{i}\mu_0[(\omega-\omega')/c^2]\{[\mathbf{J}_{\text{free}}(\mathbf{k},\omega)-\mathrm{i}\omega\mathbf{P}(\mathbf{k},\omega)]\cdot[\mathbf{J}_{\text{free}}(-\mathbf{k},\omega')-\mathrm{i}\omega'\mathbf{P}(-\mathbf{k},\omega')]\}\mathbf{k}$$
$$- \mathrm{i}\varepsilon_0\omega[k^2-(\omega'/c)^2][\mathbf{k}\cdot\mathbf{M}(-\mathbf{k},\omega')]\mathbf{M}(\mathbf{k},\omega)$$
$$+ \tfrac{1}{2}\mathrm{i}\varepsilon_0(\omega-\omega')[\mathbf{k}\cdot\mathbf{M}(\mathbf{k},\omega)][\mathbf{k}\cdot\mathbf{M}(-\mathbf{k},\omega')]\mathbf{k}$$
$$+ \tfrac{1}{2}\mathrm{i}\varepsilon_0(\omega\omega'/c^2)(\omega-\omega')[\mathbf{M}(\mathbf{k},\omega)\cdot\mathbf{M}(-\mathbf{k},\omega')]\mathbf{k}$$
$$+ (\omega\omega'^2/c^4)[\mathbf{J}_{\text{free}}(\mathbf{k},\omega)-\mathrm{i}\omega\mathbf{P}(\mathbf{k},\omega)]\times\mathbf{M}(-\mathbf{k},\omega')$$
$$- (\omega'/c)^2[\rho_{\text{free}}(\mathbf{k},\omega)-\mathrm{i}\mathbf{k}\cdot\mathbf{P}(\mathbf{k},\omega)][\mathbf{k}\times\mathbf{M}(-\mathbf{k},\omega')]$$
$$+ (\omega/c^2)[\mathbf{k}\cdot\mathbf{M}(-\mathbf{k},\omega')]\{\mathbf{k}\times[\mathbf{J}_{\text{free}}(\mathbf{k},\omega)-\mathrm{i}\omega\mathbf{P}(\mathbf{k},\omega)]\}$$
$$+ (\omega'/c^2)[\mathbf{k}\times\mathbf{M}(-\mathbf{k},\omega')]\times\{\mathbf{k}\times[\mathbf{J}_{\text{free}}(\mathbf{k},\omega)-\mathrm{i}\omega\mathbf{P}(\mathbf{k},\omega)]\}\}(\omega+\omega')$$
$$\times \exp[-\mathrm{i}(\omega+\omega')t]/\{[k^2-(\omega/c)^2][k^2-(\omega'/c)^2]\}\,\mathrm{d}\mathbf{k}\,\mathrm{d}\omega\,\mathrm{d}\omega'. \tag{B4}$$

Equations (B2) and (B4) may now be combined to yield:

$$\int_{-\infty}^{\infty}\mathbf{F}(\mathbf{r},t)\,\mathrm{d}\mathbf{r} + (\partial/\partial t)\int_{-\infty}^{\infty}(1/c^2)\mathbf{S}(\mathbf{r},t)\,\mathrm{d}\mathbf{r} = (1/2\pi)^5\iiint_{-\infty}^{\infty}\{\,k^2[\mathbf{J}_{\text{free}}(\mathbf{k},\omega)-\mathrm{i}\omega\mathbf{P}(\mathbf{k},\omega)]\times\mathbf{M}(-\mathbf{k},\omega')$$
$$-[\omega\rho_{\text{free}}(\mathbf{k},\omega)-\mathrm{i}\omega\mathbf{k}\cdot\mathbf{P}(\mathbf{k},\omega)][\mathbf{k}\times\mathbf{M}(-\mathbf{k},\omega')]+[\mathbf{k}\cdot\mathbf{M}(-\mathbf{k},\omega')]\{\mathbf{k}\times[\mathbf{J}_{\text{free}}(\mathbf{k},\omega)-\mathrm{i}\omega\mathbf{P}(\mathbf{k},\omega)]\}$$
$$+[\mathbf{k}\times\mathbf{M}(-\mathbf{k},\omega')]\times\{\mathbf{k}\times[\mathbf{J}_{\text{free}}(\mathbf{k},\omega)-\mathrm{i}\omega\mathbf{P}(\mathbf{k},\omega)]\}\}$$
$$\times (\omega'/c^2)(\omega+\omega')\exp[-\mathrm{i}(\omega+\omega')t]/\{[k^2-(\omega/c)^2][k^2-(\omega'/c)^2]\}\,\mathrm{d}\mathbf{k}\,\mathrm{d}\omega\,\mathrm{d}\omega'. \tag{B5}$$

Invoking the charge continuity condition of Eq. (9), and denoting the longitudinal and transverse components of the various vector fields in Eq. (B5) by the $\parallel$ and $\perp$ subscripts, we will have

$$\int_{-\infty}^{\infty}\mathbf{F}(\mathbf{r},t)\,\mathrm{d}\mathbf{r} + (\partial/\partial t)\int_{-\infty}^{\infty}(1/c^2)\mathbf{S}(\mathbf{r},t)\,\mathrm{d}\mathbf{r} = (1/2\pi)^5\iiint_{-\infty}^{\infty}\{\,k^2[\mathbf{J}_{\text{free}}(\mathbf{k},\omega)-\mathrm{i}\omega\mathbf{P}(\mathbf{k},\omega)]\times\mathbf{M}(-\mathbf{k},\omega')$$
$$- k^2[\mathbf{J}_{\text{free}}(\mathbf{k},\omega)-\mathrm{i}\omega\mathbf{P}(\mathbf{k},\omega)]_{\parallel}\times\mathbf{M}(-\mathbf{k},\omega')_{\perp}$$
$$+ k^2\mathbf{M}(-\mathbf{k},\omega')_{\parallel}\times[\mathbf{J}_{\text{free}}(\mathbf{k},\omega)-\mathrm{i}\omega\mathbf{P}(\mathbf{k},\omega)]_{\perp}$$
$$+ k^2\mathbf{M}(-\mathbf{k},\omega')_{\perp}\times[\mathbf{J}_{\text{free}}(\mathbf{k},\omega)-\mathrm{i}\omega\mathbf{P}(\mathbf{k},\omega)]_{\perp}\}$$
$$\times (\omega'/c^2)(\omega+\omega')\exp[-\mathrm{i}(\omega+\omega')t]/\{[k^2-(\omega/c)^2][k^2-(\omega'/c)^2]\}\,\mathrm{d}\mathbf{k}\,\mathrm{d}\omega\,\mathrm{d}\omega'. \tag{B6}$$

Finally noting that the first term in the above integrand can be expanded in terms of the longitudinal and transverse components of its constituent vectors, we find, upon reordering the cross-products

$$\int_{-\infty}^{\infty}\mathbf{F}(\mathbf{r},t)\,\mathrm{d}\mathbf{r} + (\partial/\partial t)\int_{-\infty}^{\infty}(1/c^2)\mathbf{S}(\mathbf{r},t)\,\mathrm{d}\mathbf{r} = (1/2\pi)^5\iiint_{-\infty}^{\infty}\{\,[\mathbf{J}_{\text{free}}(\mathbf{k},\omega)-\mathrm{i}\omega\mathbf{P}(\mathbf{k},\omega)]\times\mathbf{M}(-\mathbf{k},\omega')$$
$$- [\mathbf{J}_{\text{free}}(\mathbf{k},\omega)-\mathrm{i}\omega\mathbf{P}(\mathbf{k},\omega)]_{\parallel}\times\mathbf{M}(-\mathbf{k},\omega')_{\perp}$$
$$- [\mathbf{J}_{\text{free}}(\mathbf{k},\omega)-\mathrm{i}\omega\mathbf{P}(\mathbf{k},\omega)]_{\perp}\times\mathbf{M}(-\mathbf{k},\omega')_{\parallel}$$
$$- [\mathbf{J}_{\text{free}}(\mathbf{k},\omega)-\mathrm{i}\omega\mathbf{P}(\mathbf{k},\omega)]_{\perp}\times\mathbf{M}(-\mathbf{k},\omega')_{\perp}\}$$
$$\times k^2(\omega'/c^2)(\omega+\omega')\exp[-\mathrm{i}(\omega+\omega')t]/\{[k^2-(\omega/c)^2][k^2-(\omega'/c)^2]\}\,\mathrm{d}\mathbf{k}\,\mathrm{d}\omega\,\mathrm{d}\omega'. \tag{B7}$$

The integrand in Eq. (B7) is readily seen to be identically equal to zero.

**References**

1. P. Penfield and H. A. Haus, "Electrodynamics of Moving Media," MIT Press, Cambridge (1967).
2. S. R. de Groot and L. G. Suttorp, *Foundations of Electrodynamics*, North Holland, Amsterdam (1972).
3. R. Loudon, "Radiation Pressure and Momentum in Dielectrics," De Martini lecture, Fortschr. Phys. **52**, 1134-1140 (2004).
4. S. M. Barnett and R. Loudon, "On the electromagnetic force on a dielectric medium," *J. Phys. B: At. Mol. Opt. Phys.* **39**, S671-S684 (2006).
5. M. Mansuripur and A. R. Zakharian, "Maxwell's macroscopic equations, the energy-momentum postulates, and the Lorentz law of force," *Phys. Rev. E* **79**, 026608 (2009).
6. M. Mansuripur, "Radiation pressure and the linear momentum of the electromagnetic field in magnetic media," *Optics Express* **15**, 13502-18 (2007).
7. W. Shockley and R. P. James, "Try simplest cases discovery of hidden momentum forces on magnetic currents," Phys. Rev. Lett. **18**, 876-879 (1967).
8. W. Shockley, "Hidden linear momentum related to the $\boldsymbol{\alpha} \cdot \boldsymbol{E}$ term for a Dirac-electron wave packet in an electric field," Phys. Rev. Lett. **20**, 343-346 (1968).
9. L. Vaidman, "Torque and force on a magnetic dipole," Am. J. Phys. **58**, 978-983 (1990).
10. F. N. H. Robinson, "Electromagnetic stress and momentum in matter," Physics Reports (Section C of Physics Letters) **16**, 313-354 (1975).
11. D. Babson, S. P. Reynolds, R. Bjorkquist, and D. J. Griffiths, "Hidden momentum, field momentum, and electromagnetic impulse," Am. J. Phys. **77**, 826-833 (2009).
12. V. Hnizdo, "Conservation of linear and angular momentum and the interaction of a moving charge with a magnetic dipole," Am. J. Phys. **60**, 242-246 (1992).
13. M. Scalora, G. D'Aguanno, N. Mattiucci, M. J. Bloemer, M. Centini, C. Sibilia, and J. W. Haus, "Radiation pressure of light pulses and conservation of linear momentum in dispersive media," Phys. Rev. E **73**, 056604 (2006).
14. M. Mansuripur, "Electromagnetic Stress Tensor in Ponderable Media," *Optics Express* **16**, 5193-98 (2008).
15. M. Mansuripur, "Electromagnetic force and torque in ponderable media," *Optics Express* **16**, 14821-35 (2008).
16. T. B. Hansen and A. D. Yaghjian, *Plane-Wave Theory of Time-Domain Fields: Near-Field Scanning Applications*, IEEE Press, New York (1999).
17. B. U. Felderhof and H. J. Kroh, "Hydrodynamics of magnetic and dielectric fluids in interaction with the electromagnetic field," J. Chem. Phys. **110**, 7403-7411 (1999).
18. J. J. Abbott, O. Ergeneman, M. P. Kummer, A. M. Hirt, and B. J. Nelson, "Modeling magnetic torque and force for controlled manipulation of soft-magnetic bodies," IEEE Trans. Robot. **23**, 1247-1252 (2007).
19. B. A. Kemp, J. A. Kong, T. Grzegorczyk, "Reversal of wave momentum in isotropic left-handed media," Phys. Rev. A **75**, 053810 (2007).